\documentclass[a4paper,11pt]{article}
\usepackage{pos}
\usepackage{amsmath}
\usepackage{graphicx}
\usepackage{hyperref}

\title{Interpreting Reactor Antineutrino Anomalies}

\author*[a,b]{Ilham El Atmani} 
 \onbehalf{for the STEREO collaboration} 

\affiliation[a]{Institut de recherche sur les lois fondamentales de l'Univers,  CEA,  Université Paris-Saclay, \\ 91191 Gif-sur-Yvette,  France} 
\affiliation[b]{Laboratory of High Energy Physics \& Condensed Matter, Faculty of Sciences Ain-Chock,\\ Hassan II University of Casablanca,\\ B.P. 5366 Maarif, Morocco}               
\emailAdd{ilham.elatmani-etu@etu.univh2c.ma} 

\abstract{The experimental and theoretical research on the physics of massive neutrinos is based on the standard paradigm of three-neutrino mixing, which describes the oscillations of neutrino flavors measured in solar, atmospheric, and long-baseline experiments. However, several anomalies, corresponding to an \( L/E \) of 1m/MeV, could be interpreted by involving sterile neutrinos, as in the Reactor Antineutrino Anomaly (RAA) and Gallium anomaly. STEREO was designed to investigate this conjecture, which could potentially extend the Standard Model (SM). The detector provides a comprehensive study of anomalies for a pure \( ^{235}\text{U} \) antineutrino spectrum, using a Highly Enriched Uranium (HEU) core. We describe an analysis of the full set of data generated by STEREO and an accurate prediction of reactor antineutrinos. The measured antineutrino energy spectrum suggests that the anomalies originate from biases in nuclear experimental data used for predictions, while rejecting the hypothesis of a light sterile neutrino. Our result supports the neutrino content of the SM and establishes a new reference for the \( ^{235}\text{U} \) antineutrino energy spectrum.}

\FullConference{42nd International Conference on High Energy Physics (ICHEP2024)\\
18-24 July 2024\\
Prague, Czech Republic\\}

\begin{document}
\maketitle

\renewcommand{\figurename}{Figure}

\section{Introduction}
Reactor neutrinos are produced by the decay of fission products. The Reactor Antineutrino Anomaly (RAA) \cite{Mention,Mueller} shows a 6.5\% deficit in measured flux with respect to prediction at short baselines from reactor cores, inconsistent with the three-flavor oscillation model and suggesting a potential sterile neutrino state. Additionally, the shape anomaly indicates a 10\% local excess in the neutrino energy spectrum, centered around 5 MeV, representing only 1\% of the overall
flux. The STEREO experiment aims to address these anomalies with a precise and model independent measurement of the neutrino fission spectrum.

\section{STEREO experiment}

The STEREO detector was installed 10 m from the compact core of the ILL reactor, in Grenoble, France, with a thermal power of 58 MW. Due to the highly enriched fuel (93\%) the measured neutrino spectrum is purely from \( ^{235}\text{U} \) fissions, with negligible contribution from \(^{239}\text{Pu}\) fissions. The target volume was filled with Gd-loaded liquid scintillator. It was segmented in six identical cells allowing an oscillation analysis based on the relative comparison between cells, independent of any spectrum prediction. Thanks to a combination of passive and active shielding, overburden from of water channel above the experiment, and high discrimination power of the IBD detection process, STEREO achieved a signal-to-noise ratio close to 1.\\
Regular calibrations with a large set of radioactive sources circulated inside and around the target volume ensured an accurate control of the energy scale. Reconstruction of the beta-spectrum of the cosmogenic \( ^{12}\text{B} \) further refined the energy scale through global fitting, with all residuals contains within \( \pm 1\% \). The precision of the absolute normalization is driven by the neutron detection efficiency (\( \varepsilon_n \)) and the thermal power of the reactor (\( P_{\text{th}} \)) . An unprecedented uncertainty of 0.7\% was obtained for \( \varepsilon_n \) thanks to the use of the FIFRELIN code to describe the de-excitation \( \gamma \)-cascades from neutron capture on Gd. A careful study of the instrumentation in the primary loop allowed to reach a 1.4\% uncertainty on \( P_{\text{th}} \), which is quite satisfactory for a research reactor.\\
STEREO has collected 107 k \( \nu_e \) over 273 days of reactor operation (7 cycles) \cite{Almazan}. The dominant cosmic-rays induced background was accurately subtracted using interleaved periods of reactor OFF (520 days total) and sub-\% level monitoring the detector response across ON and OFF periods.  

\section{Results}
\begin{figure}[h]
   \centering
    \begin{minipage}{0.52\textwidth}
        \centering
        \includegraphics[width=\linewidth]{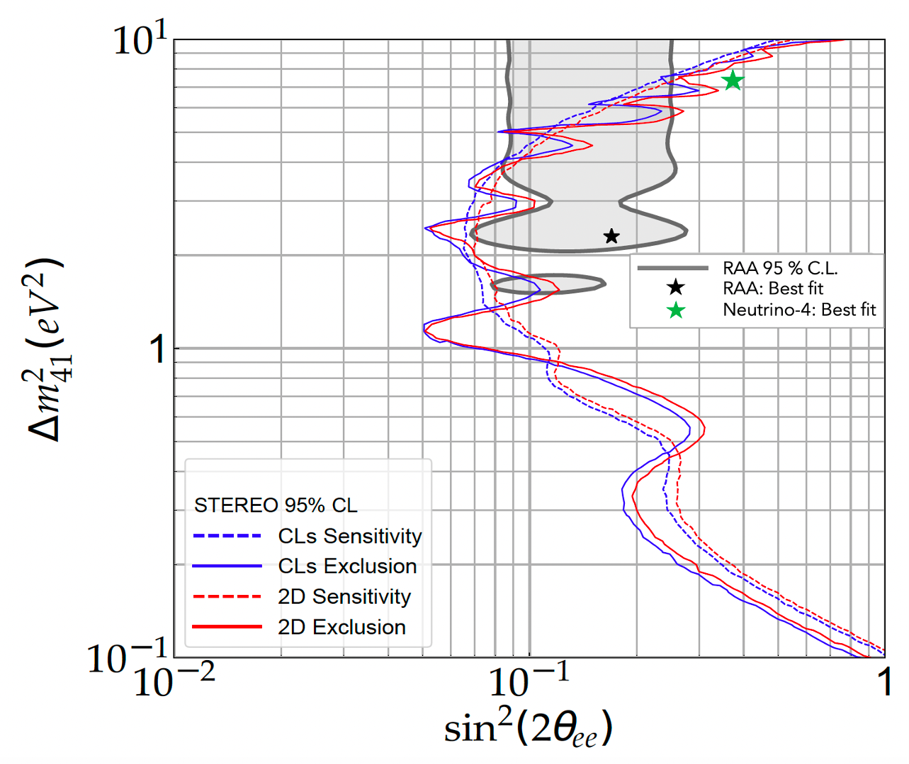} \\
        \caption{Exclusion contours (solid red line) and exclusion sensitivity contours (dashed red line) of the STEREO experiment at 95\% CL. These results are derived in a 2D Feldman–Cousins framework (red) and verified by following the Gaussian CLs alternative statistical prescription (blue). The grey contour shows the initial RAA contour..
 } 
         \label{figure1:example1}
    \end{minipage}%
    \hfill
    \begin{minipage}{0.45\textwidth}
        \centering
        \includegraphics[width=\linewidth]{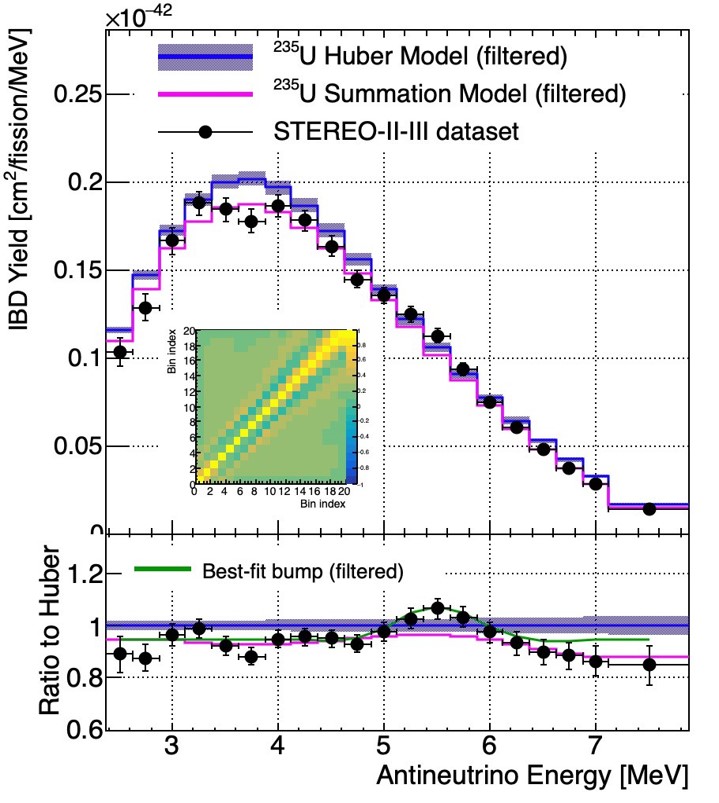}
        \caption{The unfolded antineutrino spectrum associated to the fission of \(^{235}\text{U}\) (black points) is shown with the Huber model (blue) and summation model predictions (purple) in the true antineutrino energy space.
 }
         \label{figure2:example2}
    \end{minipage}
\end{figure}

The oscillation analysis is performed by fitting a free mean neutrino spectrum to the spectra measured in the six identical cells. The existence of an oscillation pattern is then tested by comparing the cell-to-cell relative distortions to this common spectrum with the distortions expected from a given set of oscillations parameters (mixing angle and frequency). The results shown in (Figure~\ref{figure1:example1}) are derived in a 2D-Feldman-Cousins framework. The data are found compatible with the no-oscillation hypothesis. The associated exclusion contour rejects with high confidence level (CL) most of the sterile \( \nu \) parameter space expected from the RAA. The best-fit point of the Neutrino-4 experiment is also excluded at \(3.3\,\sigma\).\\
By combining the compatible spectra from the six cells a reference \( ^{235}\text{U} \) fission neutrino spectrum is obtained. Based on the very good control of the detector response demonstrated by the \%-level agreement between various calibration data and simulation, a response matrix of the detector is build and the measured neutrino spectrum is corrected for all detection effects using a regularized Tikhonov approach \cite{Almazan}. The spectrum in “true” neutrino energy space is shown in (Figure~\ref{figure2:example2}). Its comparison with the Huber Model (HM) confirms the mean deficit of the RAA and a local distortion fitted by a Gaussian function centered on 5.5 MeV with  \(15.6 \pm 5.2\%\) amplitude.

\section{Conclusions}

STEREO offers a reference measurement of the \(^{235}\text{U}\) neutrino spectrum. It confirms the deviation of the “5 MeV bump" from the predicted spectrum shape and the \(5.5 \pm 2.1\%\) rate deficit. The hypothesis of an oscillation toward a sterile neutrino state with mass around 1 eV is rejected with high confidence level. These results set a new benchmark for nuclear data and point to point to a bias in the nuclear data used for the normalization of the predictions as the most probable explanation of the RAA. Joint analyses with Daya Bay and PROSPECT  experiments are ongoing to improve further the accuracy of reference neutrino spectra.

\end{document}